# Applying Convolutional Neural Networks for Stock Market Trends Identification

[0000-0002-1516-7378] Ekaterina Zolotareva

[1]Data Analysis and Machine Learning Department, Financial University under the Government of the Russian Federation, 38 Shcherbakovskaya St., Moscow 105187, Russia

`elzolotareva@fa.ru`

**Abstract.** In this paper we apply a specific type ANNs - convolutional neural networks (CNNs) - to the problem of finding start and endpoints of trends, which are the optimal points for entering and leaving the market. We aim to explore long-term trends, which last several months, not days. The key distinction of our model is that its labels are fully based on expert opinion data. Despite the various models based solely on stock price data, some market experts still argue that traders are able to see hidden opportunities. The labelling was done via the GUI interface, which means that the experts worked directly with images, not numerical data. This fact makes CNN the natural choice of algorithm. The proposed framework requires the sequential interaction of three CNN submodels, which identify the presence of a changepoint in a window, locate it and finally recognize the type of new tendency - upward, downward or flat. These submodels have certain pitfalls, therefore the calibration of their hyperparameters is the main direction of further research. The research addresses such issues as imbalanced datasets and contradicting labels, as well as the need for specific quality metrics to keep up with practical applicability. This paper is the full text of the research, presented at the 20th International Conference on Artificial Intelligence and Soft Computing Web System (ICAISC 2021)

## 1   Introduction

An ability to identify stock market trends has obvious advantages for investors. Buying stock on an upward trend (as well as selling it in case of downward movement) results in profit, which makes predicting stock markets a highly attractive topic both for investors and researchers. Despite the long history, this field, as stated in [1] is still a promising area of research mostly because of the arising opportunities of artificial intelligence.

Modern machine learning technologies are presented by a number of various algorithms, but in general, there are three main classes of models used for the prediction of stock markets - artificial neural networks (ANNs), support vector machines (SVM/SVRs) and various decision tree ensembles (e.g., Random forests). As of 2017, the hegemony of ANNs and SVM/SVRs has been observed - the articles based on these models accounted for 86% of articles researched in [1]. Among twenty scientific papers on stock market forecasting published in the period between 2018 and early 2021, at least eight studies [2–9] exploited ANNs as the only main algorithm, another two used ANN in the ensemble (or stack) with other algorithms of equal importance [10, 11]. In four studies ANNs were used to compare performance with other algorithms, chosen as main [12–15].

It should be noted, that despite the common main method, different researches have sufficient variations in model structure, problem formalization, and features and labels accordingly. Models also may be applied to different markets, assets and prediction horizons. A certain variation also exists in the selection of performance measures which should ensure the comparability of models.

The absolute majority of researches, which apply ANNs as the main algorithm (or part of an ensemble or stack) to stock market forecasting [2, 3, 6–10] are concentrated on predicting the direction of the stock market, thereby solving a classification problem ("direction type" studies). Fewer predict prices [4, 11, 15] 11, 15] by solving a regression problem ("price type" studies). In all cases the ground truth variable is extracted from the historical price series.

All the "price type" studies and most of the "direction type" studies focus on daily basis predictions and only in three papers the time horizon varies from one week [2, 7] to one month [10].

Another important difference between the suggested models is the choice of features. Though all of the researches used market data (e.g., prices, volumes and the values derived from them - technical analysis indicators, correlations, volatilities and returns) as input variables, in some studies they were supplemented by text features [7, 11] or fundamental macroeconomic variables [6, 10].

In this paper we apply a specific type of ANNs - convolutional neural networks (CNNs) - to the problem of finding start and endpoints of trends. CNNs have appeared and developed largely due to the increased need in solving computer vision problems. In 2011 the AlexNet convolutional network [16] led to a breakthrough in the field of image classification. Subsequently, CNNs were used not only in the problem of classification, but also in image detection (for example, the YOLO methodology [17], as well as the RCNN, Fast-RCNN, and Faster-RCNN algorithms [18]), therefore their application to stock market forecasting is less common compared to the traditional fully connected networks. Here we consider a mathematical model based on a convolutional neural network, which is used not for the classical problem of image classification and detection, but for recognizing the state of the financial market (upward trend, downward trend, or flat) and predicting future moments of a trend reversal.

Among the reviewed recent papers on stock market forecasting, only three have applied CNNs. [2] suggests an integrated framework, which fuses market and trading information for price movement prediction. CNN is used to extract trading features from the transaction number matrix, the buying volume matrix, and the selling volume matrix of investors, clustered by their trading behaviour profile. The output of CNN is concatenated with the market information weighted by stock correlation and then is fed into another deep neural network algorithm to obtain predictions for the several following trading days.

[6] apply CNN to capture correlations among different variables for extracting combined features from a diverse set of input data from five major U.S. stock market indices, as well as fundamental macroeconomic variables (e.g., currency rates, commodity prices, etc.) to predict next-day price movements.

[13] address the problem of stock preselection for portfolio optimization. The authors consider the performance of five machine learning algorithms - random forest, support vector regression (SVR) and three neural networks (LSTM neural network, deep multilayer perceptron and CNN) - inputting past 60 days' daily returns to predict the next day's return.

Unlike the majority of other researchers, we aim to explore long-term trends, which last several months, not days. The start and endpoints of such trends accordingly are the optimal points for entering and leaving the market. Despite the various technical analysis algorithms and econometrics studies based solely on stock data, some market experts still argue that traders are able to see opportunities of making money (i.e., detecting trends or turning points) that cannot be formally expressed. Thus, using computer science algorithms to learn from successful traders' decisions (and not only stock data) is likely to improve financial market models.

The key distinction of our model is that its ground truth vector is fully based on expert opinion data, provided by one major investment company. The basis for building a mathematical model is the historical data of the financial market, divided by experts into markup windows, each of which corresponds to some unchanged market state. Unlike other researchers, we do not use a mathematical formula to define a trend, instead, it is defined by an expect as a potentially profitable (or unprofitable) pattern in price dynamics. The major difficulty of this approach is that it is exposed to subjective judgements of the experts. On the other hand, if the experts are successful traders in a certain investment company, it gives the employer the chance to 'digitalize' their exceptional skills and obtain a machine learning algorithm no one else on the market can employ.

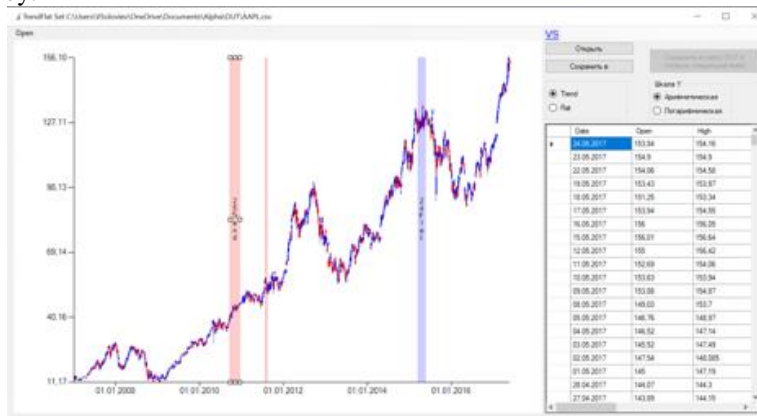

**Fig. 1.** GUI interface for data labelling

The labelling of market states was performed in a specially designed graphical interface, where the experts marked certain consequent periods as "Trend", "Flat" or N/A (see

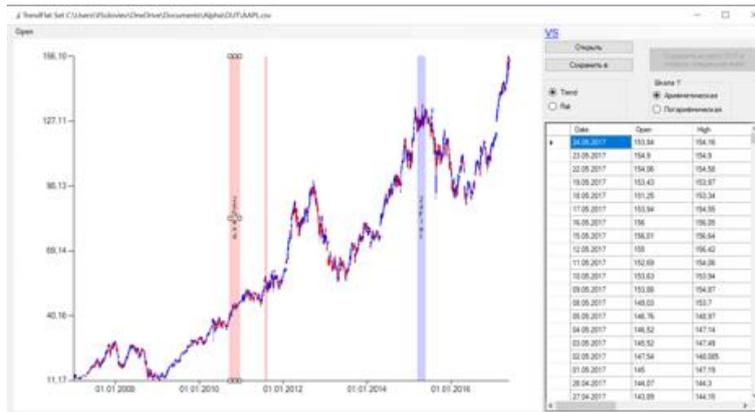

). They worked directly with images, not numerical data. This fact makes CNN the natural choice for the case, as we need to extract implicit patterns from images which is exactly the original mission of CNN.

The model requires only raw historical price data as input. From one point of view, it can be considered a limitation, since we ignore fundamental factors and news feed. On the other hand, it makes the model unpretentious in production, since stock data is easily obtainable and can be downloaded into the company's informational systems or directly fed into the model via API. Besides, the concept of using prices as only input features corresponds with the hypothesis of market efficiency (EMH) in the sense that prices reflect all the available market information [19].

Technically we are solving a classification problem (it is a "direction type" research), but the standard quality metrics (accuracy, AUC and F1-Score) turn out to be inapplicable because of imbalanced datasets and contradicting labels. For these reasons, the returns of simulated trading are used as the main performance indicator.

## 2 The Proposed Framework

The initial markup contained three types of windows - trend, flat or unknown state. Approximately 90% of identified trends and flats last from 40 to 600 business days, which is in line with medium- and long-term trends. Initially, the task was set as recognizing the point of a change in the market state (the transition from trend to flat and vice versa, regardless of the trend direction) with a minimum time lag - for example, identifying the beginning of a 200-day trend 20 days after its start. However, as the study progressed, it also became necessary to distinguish between the direction of trend (upward and downward) to be able to assess the returns and compare various modifications of the model.

The practical purposes of the study impose several restrictions. First, the constructed model must be universal, that is, it must not be limited to a specific asset, market or time period. After training and calibration, it should be applicable for any financial instrument for which the standard price data (daily open and close prices, high, low) and trading volumes are available. Another significant issue, which should be considered when working with historical data, is the absence of future prices and volumes at the time of calculations. Violation of this condition, as well as a significant increase in time lag in changepoint identification, makes the simulation results inapplicable in practice.

### 2.1   Model structure

**The Submodels.** The proposed framework for predicting future changepoints in the market state includes three components.

- ChangePoints_classifier (abbreviated to ChP-c), the classification model that determines the presence or absence of the changepoint at a certain time interval. For the output the model returns the value 1 ("The trend changed at the specified interval") and 0 ("The trend did not change at the specified interval").
- ChangePoints_regression (abbreviated to ChP-r), the regression model that determines the position of the last changepoint at a certain time interval, provided that at least one change of state has occurred. The position is determined by a real number in the range from 0 to 1 (proportion of the width of the time interval), where 0 means that the last changepoint was recorded at the beginning of the period, and 1 - at the end. Knowing the start date and the width of the time interval, we can convert the value obtained from the ChP-r model to a specific date.
- Trend_or_Flat (abbreviated as TF), the classification model that determines the type of trend observed in a given markup window. For the output the model returns the value 1 ("Upward trend"), -1 ("Downward trend") or 0 ("No trend, flat").

It was mentioned above that one of the practical requirements for the model is to ensure its universality (that is, the independence of a specific financial instrument, and, in particular, of the range of changes in its prices or trading volumes in any time interval). It is also unacceptable to "look into the future", which makes it impossible to normalize quantitative indicators with, for example, annual or monthly highs and lows - they are unknown at the time of calculation. One of the possible options for constructing a mathematical model under such conditions is to interpret the quotes of a financial instrument not as a set of quantitative indicators, but as an image, for example, a chart with Japanese candlesticks and corresponding labels. Image detection assumes recognition of both the type of object (classification problem) and determination of the boundaries of its location on a snapshot containing several types of objects of different sizes (regression problem). In our case, the problem of detection can be interpreted as determining the type of trend and its boundaries (beginning and end) in the time window which plays the role of a snapshot. Thus, the input data in all three submodels (ChP-c, ChP-r,

TF) are matrices of digitalized images corresponding to the quotes charts on certain time intervals - data slices.

**The Interaction of Submodels**. The interaction of submodels in real-time simulation (that is when new information about quotes arrives every day) is shown in **Fig. 2**. The ChP-c model, which is responsible for determining the presence of changepoints in a slice with a width of n working days (for example, n_days = 25 corresponds to approximately 5 weeks, and n_days = 75 to 15 weeks), is fed a digital image - a quote chart for the selected data slice. To reduce the computation time, data slices can be taken not every day, but with a skip step (for example, skip = 5 days), which gives a non-critical error. If the ChP-c model detects the presence of a changepoint in the window (value 1), then the ChP-r model is activated, which determines exactly where in this data slice the last trend started (win_srt). If the ChP-c model returns the value 0 (there was no trend change), then the trend start point (win_srt) is set to the value determined from the previous slice (if it is absent, then win_srt is equal to the beginning of the quotes history). By default, the end of the trend (win_end) is the current date on which the ChP-c model is launched. A digital chart of stock quotes taken for the dates between win_srt and win_end is transmitted to the TF model, which determines the type of trend (uptrend/downtrend or flat) and the corresponding recommendation - to open and hold a long/short position or close the position.

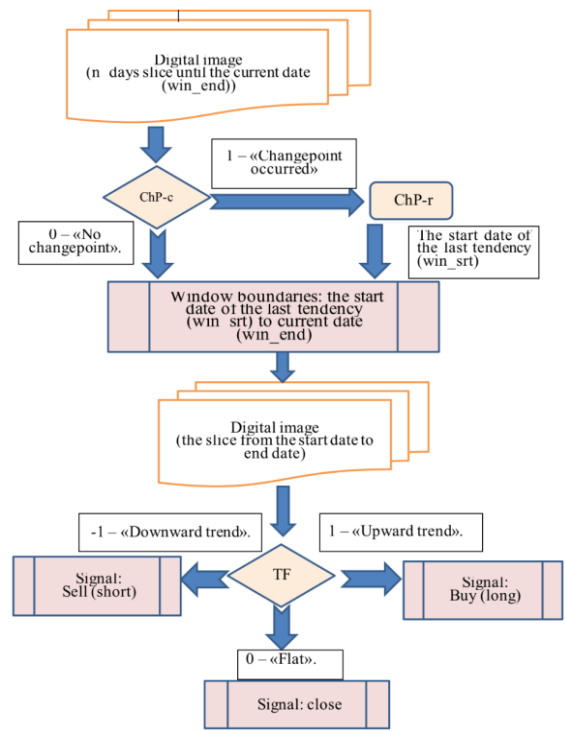

7**Fig. 2.** The interaction of submodels.

## 2.2 Data Preparation

The dataset to explore consists of 1389 files labelled by 2 experts. The data contains quotes of 700 stocks included in the S&P index, covering the period from 2005-01-28 to 2017-09-13. Each file contains on average around 2600 daily quotes (Date, Open, High, Low, Close) for a certain period and stockname, labelled by a certain expert.

**Image Processing.** As noted above, the input data in all three submodels (ChP-c, ChP-r, TF) are matrices of digital images corresponding to the quotes charts on certain time intervals - data slices.

One of the most common chart types that accounts for all price data (open price, close price, high and low) is the Japanese candlestick chart. In this case, the image turns out to be coloured, where the green colour of the candlestick body, as a rule, corresponds to the increase in price (the closing price is higher than the opening price), and red - to the decrease. The choice of colours, however, is not critical, as long as they are contrasting.

From the point of view of digitization, an image is a collection of pixels arranged in a given order. Each pixel is characterized by its intensity, measured in the range from 0 to 255 (from white to black for monochrome). To digitize colour images the RGB scheme is often used. In this case, each pixel is described by a three-dimensional vector containing the colour intensity in three channels - red, green and blue. All other colours are obtained by combinations of colour intensities in different channels.

The intensity value in each channel of a pixel corresponds to one variable, therefore, the total number of input variables depends on the image resolution (dpi, the number of digits per inch) and the number of channels describing one pixel. The decrease in the resolution and the number of channels (and, as a consequence, a decrease in the dimension of the feature space) increases the speed of calculations but can lead to the loss of some information.

**Logarithms.** The initial data contains the actual values of price and volume quotes. At the same time, when considering the dynamics of financial markets over long time intervals, it is often preferable to use a logarithmic scale, since it is less sensitive to the spread of absolute values. When marking up, the experts used the logarithmic scale, which gives a slightly different visual effect, therefore, in this study, all the quotes, which served as the basis for the input variables, were replaced by their natural logarithms.

**Duplicates, Contradictions and Imbalanced Datasets.** Different experts could have marked the same "data points", but their markup does not always coincide. In our case, the number of duplicates with the same "Type" field value is 1,246,726 (37.8%) "data points". The opposite situation (expert opinions for the same "data point" does not coincide) leads to contradictions. The number of such records is 397,321, which is approximately 24% of 1,660,441 - the number of unique "data points".

There are at least three options for working with such data:

1. use the original data array, without removing duplicates (expert opinions for the same data point coincide) and contradictions (expert opinions for the same data point do not coincide);
2. use the original data array, in which duplicates are excluded, while the contradictions remain. This gives an obvious advantage due to the exclusion of the repeated part of the sample from the calculations and an increase in calculations. At the same time, the presence of duplicates can be viewed as an indicator of expert confidence in assessing market characteristics, which can be an important signal for training the model. Eliminating duplicates eliminates this hidden dependency;
3. use the original data array, from which both duplicates and contradictions are excluded. The contradictions undoubtedly complicate the assessment of the model, since all the main quality metrics (Accuracy, AUC, Precision, Recall, F-Score for classification problems, determination coefficient $R^2$ for regression) are based on comparing model predictions with labels given that the same combination of input variables corresponds to the only correct value of the label ("ground truth"). From this point of view, contradictions should be excluded from the sample. On the other hand, the presence of contradictions, as well as in the case of duplicates, can be interpreted as an indicator of experts' confidence in assessing market characteristics. By eliminating contradictions, we risk losing this important information. In this regard, the strategy for dealing with contradictions should be based on the study of the causes of their appearance, which are the following (we assume the high qualification of experts):
   a. The experts are different traders. One would see a long upward trend, the other would see two middle-term trends interrupted by a short downward or flat period. Where the first sees no changepoints, the other would see two.
   b. Since the labelling was done in a GUI by mouse-clicking, technical blots might have occurred - one might fail to select the precise datapoint because of improper screen settings or poor eyesight.

It should be noted that the problem of duplicates and contradictions is more significant for the models that determine the changepoints (ChP-c and ChP-r) than for the TF model: almost for every "data point" labelled as changepoint, there exists a "no changepoint" label obtained from another expert. Within the framework of this study, we corrected only the technical blots, which eliminated 8% of contradictions. The possibility of improving models by removing duplicates and contradictions (for example by averaging the experts' opinions or voting) is the subject of further research.

Another issue is that the dataset is imbalanced. In the ChP-c model, most observations do not contain changepoints and the imbalance increases from 3:1 to 10:1 as n_days, the width of the data slice, decreases from 75 to 25. This is quite understandable since the changes in long-term and medium-term trends change occur once every several hundred working days. As for the TF model, more than half of the windows in both train and test set are labelled as flat, and there are about 2 times more upward trends than downward.

Ideally, the proportion of classes in the binary classification model should be close to 1:1, since as the imbalance increases, observations of the majority class begin to "suppress" observations of the minority class. To deal with imbalanced classes one can use oversampling or undersampling techniques, which alter the original dataset. Another approach would be to modify the loss function or the learning algorithm itself so that it becomes more responsive to observations of the minority class. In this study, the choice was made in favour of modifying the standard loss function.

**Train and test samples**. To assess the generalizing ability of the model, the data array is traditionally divided into training and test samples. These samples must be independent, otherwise, the quality metrics will be unduly overestimated. We have considered three possible schemes for dividing the sample into training and test in a more or less standard proportion of 70% by 30%:

1. random partition – is deemed inapplicable. The presence of duplicates in both the training and test samples will violate the independence condition. One of the solutions is to manually exclude duplicate records from the sample, but since the ChP-c, ChP-r and TF models use different sets of images for training, it will be difficult to completely eliminate the intersection of samples;
2. division by assets. The data array is represented by 700 stocks, which could also be divided into training and test samples in the required proportion. However, the correlation can arise due to overlapping time intervals. Since quotes are taken from the same market, different assets will implicitly repeat the general market trends characteristic of a certain time period. This would also violate the requirements for the independence of the training and test samples, although less obvious;
3. division by date. To ensure data independence, the final solution is to use 70% of the earlier quotes as a training sample, and the remaining 30% as a test sample. The threshold date separating the test and training samples is set to October 17, 2014.

## 3 The GENERAL SCHEME FOR CNN

The general scheme for building a CNN includes several stages. At each stage, it is necessary to fix some parameters (more precisely, hyperparameters), on which the result of training will largely depend. The actual machine learning algorithm is implemented in Python 3.5/3.6 using the CNTK library from Microsoft (v.2.5.1) [20], Pandas 0.22.0, Numpy 1.14.2 and Matplotlib 2.2.2. A similar solution can also be implemented in the libraries neural network TensorFlow, Caffe, etc., but will require adaptation considering the specifics of their syntax, requirements for data formats and compatibility with software platforms.

### 3.1 Extracting Features and Labels

At this stage, it is necessary to fix the following data preprocessing settings:

— image characteristics: resolution, number of channels. In the present study, colour images with a resolution of dpi = 10 were used to train the ChP-c and ChP-r models.

For the TF model, which had a smaller sample size, the 20 dpi and 60 dpi options were also tested. It did not reveal any significant quality benefits but resulted in increased data volume and processing time. Replacing colour images with black and white is the subject of further research since, obviously, it will reduce the resource intensity and computation time.
- the procedure for dealing with duplicates and contradictions. Possible options, their advantages and disadvantages are described in section 2.2.3.
- the width of the data slice (n_days) and the step (skip) with which they are taken - for the models that determine the changepoints (ChP-c and ChP-r). In the designed models, the options n_days = 25 and n_days = 75 were used, which is slightly longer than monthly and quarterly intervals. The skip step is actually a technical hyperparameter that limits the size of the resulting sample (reaches several gigabytes) and, accordingly, the time of its generation (can take tens of hours).

The generated sets of features and labels are saved in a special text format CTF, compatible with the features of machine learning procedures for models using the CNTK library.

An example of data recording in CTF format for TF model is shown in **Fig. 3** (each line corresponds to one observation):

```
|labels 0 1 0|features 255 0 0 255 ….255
|labels 0 0 1|features_0 0 0 255 … 0
….
```

**Fig. 3.** An example of data in CTF format

The size of such a file, containing about 100 thousand lines with 9216 thousand variables (one colour image with dpi = 10), is about 3-4 GB, however, the physical volume can be halved if we switch to the colour scheme containing only the values 0 and 255. In this case, they can be normalized by dividing by 255 and stored as 0 and 1.

### 3.2 CNN Structure

The convolutional neural network model can have three types of layers, characterized by different sets of hyperparameters. A multidimensional array containing input variables (features) is fed as the input of the neural network. For the output the network returns labels.

1. Convolutional layers (required). Each convolutional layer is characterized by the following hyperparameters:

- Activation function. In all the constructed models, the ReLU function was used as the activation function of the convolutional layer, which is considered the best option for the operation of training algorithms for convolutional networks.
- Weights initialization method. The main requirement for the weights initialization method is that their initial values are chosen randomly. In the framework of this study, when initializing the weights, we mainly used the modification of the uniform distribution (glorot.uniform in the CNTK library). Attempts to use other modifications of it, as well as modifications of the normal distribution, did not have any significant effect on the result.
— Filter size (f), number of filters (n), stride(s), padding (p). The output of each previous intermediate layer is the input for the next. The parameters f, n, s and p determine the sizes of the output arrays on each of the layers, as well as the number of parameters (weights), the values of which are adjusted in the learning process. It is obvious that an increase in the dimension of arrays, as well as an increase in the number of parameters, leads to an increase in the resource intensity of the learning process. All the convolutional layers used padding with automatic selection of the size p, so that the width and height of the input array, provided that a stride $s = 1$ was used, had the same padding.
— The number of convolutional layers. Most of the built models used 2 or 3 convolutional layers. At the same time, it can be noted that an increase in the number of layers did not have a cardinal effect on the result.

2. Pooling layers (optional). They are characterized by the following hyperparameters:

— Pooling layer type. Possible options - max pooling, average pooling.
— Filter size (f), stride (s), the number and the location of the pooling layers.

Pooling layers, alternating with convolutional layers, are designed to reduce the dimension of the arrays passed to the next layer of the network. The size of the filter and the stride determines how much the dimension is reduced. In the present study, adding a pooling layer did not significantly affect the speed and quality of training. This is probably due to the relatively small size of matrices.

3. Fully connected layers (as a rule, they are the last and are necessary to obtain the output variable in the form of a number or vector). They are characterized by the following hyperparameters:

— Size (number of neurons). If the layer is the last, then the number of neurons in it must match the dimension of the label space. For binary classification problems (ChP-c model) and regression (ChP-r model), the label is represented by a scalar, for three-class classification (TF model) the labels are represented as a one-hot-encoding vector, the length of which is equal to the number of classes, that is, three.
— Activation function. For the regression problem, a linear activation function is usually used, for a binary classification - a sigmoid one (this provides values in the range from 0 to 1). For multiclass classification with labels in the form of a one-hot-encoding vector, a linear activation function can also be used, which is fed to the input to the softmax operator (returns a vector, the sum of the values of which is equal to one)
— Weights initialization method. Similar to convolutional layers. The main requirement for the way weights are initialized is that their initial values are randomly selected.

‒ Number and location of fully connected layers. In the framework of this study, only one fully connected layer was used (the last one). The experiment with the addition of an extra fully connected layer did not have a significant effect on the result.

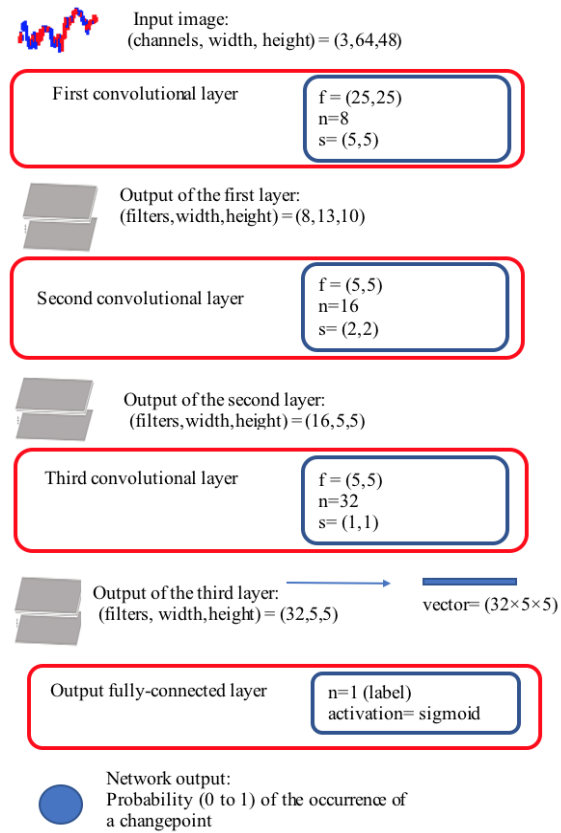

**Fig. 4.** The final structure of the ChP-c model.

**Fig. 4** presents the final structure of the ChP-c model. The total number of parameters (weights) of the model in eight blocks is 32 117. The ChP-r structure is similar except for the output layer – it uses linear activation function instead of sigmoid. TF models have two convolutional layers instead of three (the parameters f (filter size), n (number of filters), s (stride) depend on the original resolution of images –**Table 1**) and, of course, a different fully-connected layer with a vector output which is fed into softmax operator.

**Table 1.** The structure of the two best TF models.

| dpi | 1st conv | 2nd conv |
|---|---|---|
| 10 | f=(25,25) | f=(5,5) |
|    | n=8 | n=16 |
|    | s=(5,5) | s=(2,2) |
| 20 | f=(50,50) | f=(10,10) |
|    | n=8 | n=16 |
|    | s=(10,10) | s=(4,4) |

### 3.3 CNN Training

The general logic of the machine learning process, regardless of the class of models used, is based on minimizing a certain loss function that estimates the degree of deviation of model predictions from labels for various values of parameters (weights). The optimal set of the parameters (weights) which correspond to the minimum of the loss function are the main result of the training process. The minimization of the loss function is carried out by numerical methods (for example, using gradient descent which ensures, at each iteration, a sequential approaching to the minimum with a certain learning rate $\alpha$. The technical implementation of the learning process determines its resource intensity. In some algorithms, in order to increase the speed of calculations on large amounts of data, the calculation of the loss function on each iteration is performed on a random subsample (of a smaller size) of the entire training sample. Such subsamples are called minibatches.

The process of training a convolutional neural network in the CNTK library requires the definition of the following hyperparameters:

– Loss Function Type. The loss function type depends on the machine learning problem. We used the weighted binary cross-entropy function and the f-measure function (based on the F-score metric) for the problem of binary classification (ChP-c model), the squared error function for the regression problem (ChP-r model) and the cross-entropy function with softmax operator for three-class classification (TF model). Experiments with various options of the loss function, and, in particular, adding weighting, have shown that the choice of learning function has a crucial impact on the learning outcome.
– Optimization algorithm. Almost in all the constructed models, the stochastic gradient descent algorithm with subsampling minibatch was used. Experiments with alternatives (Adam, Adagrad) did not reveal any benefits.
– Minibatch size. The minimum size of the minibatch can be equal to one observation, but in this case the loss function behaviour will be unstable. At the same time, the large size of the minibatch will make the learning process unnecessarily resource-intensive. In the present study, minibatches of 64 or 8 observations were mainly used depending on the size of the training sample and/or the required number of iterations.
– The number of iterations. At each iteration, the loss function is recalculated based on the minibatch subsample and the weights are corrected to bring the loss function closer to the minimum. With a fixed minibatch size, the number of iterations can be determined by the size of the training sample and the number of its runs. There is no strict requirement regarding the optimal number of iterations. Ideally, training should stop when the value of the loss function stabilizes - that is when the next iteration does not bring us closer to the minimum.

— Learning rate α. The learning rate is responsible for the degree of weight correction at each iteration of the optimization algorithm. Choosing the too small value of α leads to a slowdown in the learning process, but too large α can lead to skipping the minimum of the loss function, and the model also stops learning. One of the machine learning heuristics suggests that α = 0.001 is optimal for computer vision models[1]. The results of this study are consistent with this heuristic. The experiments with increasing and decreasing the value of α by a factor of 100 led to a significant deterioration in learning outcomes.. The results of this study are consistent with this heuristic. The experiments with increasing and decreasing the value of α by a factor of 100 led to a significant deterioration in learning outcomes.

The choice of model hyperparameters can have a decisive influence on the simulation result, however, the problem of finding the optimal combination of hyperparameters is nontrivial and requires a series of experiments. For simple models, automated procedures are suitable - full or randomized search on a grid composed of all possible combinations of hyperparameters. For more complex models, grid search is usually associated with technical difficulties, since running one cycle can take several hours or even days. In such situations, heuristics obtained as a result of similar studies and logical analysis of pitfalls of each experiment play an important role. Within the framework of this study, it was revealed that the choice of the type of the loss function and the learning rate α had a decisive influence on the learning outcome. Variations of other parameters either did not have any significant effect or influenced only the duration and resource intensity of the learning process (if the model quality is sufficient enough, the preference, of course, was given to faster and less resource-intensive combinations of hyperparameters).

### 3.4 CNN Validation

The CNN validation procedure is similar to other machine learning models and is determined by the type of problem. For the classification problem, the main quality metrics are Accuracy and the Confusion matrix (ChP-c, TF models), as well as indicators AUC, Precision, Recall, F-Score in the case of binary classification (ChP-c model). For the regression problem (ChP-r model), the coefficient of determination $R^2$ and the mean absolute error (MAE) are usually used.

Nevertheless, it should be kept in mind, that the final prediction of the model is formed as a result of the interaction of all three submodels. In this regard, the quality metrics calculated for each of the submodels separately can serve only as one of the indicators but are not decisive. Another reason why standard quality metrics are not quite applicable for the purposes of this study is the time factor. In a common classification problem, observations are assumed to be independent, so shifting all the predictions by one record forward is likely to result in an error. At the same time, when working with time series, even if the data is taken with a skip = 5 days, a shift in predictions means only a slight delay in the model, and, therefore, a slight decrease in the financial result. This leads to the conclusion that in our case, from the practical point of view, the main quality criterion is the returns obtained as a result of the interaction of all the three submodels (ChP-c, ChP-r and TF).

---

[1] https://stackoverflow.com/questions/41488279/neural-network-always-predicts-the-same-class).

## 4    The RESULTS

### 4.1    Submodel Results

**ChangePoints_classifier**. The train and test data set for the ChP-c models have around 90 thousand and 40 thousand of observations accordingly (the exact number depends on n_days and skip parameters). In total 24 more or less successful ChP-c models were trained (we do not consider experiments with the absolutely unfortunate choice of hyperparameters). They differ in the n_days (25 and 75) parameter, dataset choice (labelled by both expert or only one to exclude contradictions), loss function modification and the number of iterations. The quality metrics of the two best models, calculated on the test set, are summarized in **Table 2**. The F-score in parentheses contains the values of the F-score metric for the minority class - F-score$_{y=1}$.

**Table 2.** The best ChP-c models.

| Model id | ChP-c_4 (n_days=25) | ChP-c_6 (n_days=75) |
|---|---|---|
| Accuracy | 81.92% | 66.42% |
| AUC | 54.90% | 56.43% |
| F-score | 83% (24%) | 64% (28%) |
| False negatives | 85% | 76% |
| False positives | 11% | 17% |

Despite all the efforts made to deal with the imbalanced dataset, both models have the same pitfall: they often miss real changepoints (give false negative predictions) - that is they ignore the minority class. Let us keep in mind, however, that because of the contradictions issue, the metrics based on ground truth may be inaccurate.

**ChangePoints_regression**. In total 19 ChP-r models were trained. They differ in the n_days parameter, skip parameter, dataset choice, loss function modification, number of iterations, number of layers and other hyperparameters. The values of the n_days parameter were chosen in such a way as to ensure the compatibility with the corresponding ChP-c models. Since we now deal only with the windows where changepoint occurs, the total number of observations is certainly less than in the ChP-c model with skip =25. However, the sample size can be adjusted by reducing the skip to 2 or 5 days to give around 100 thousand and 50 thousand observations on train and test accordingly. The quality metrics of the two best models on the test set are summarized in **Table 3**.

Table 3. The best ChP-r models.

| Model id | ChP-r_13 (n_days=25) | ChP-r_15 (n_days=75) |
|---|---|---|
| $R^2$ | 0.133 | 0.112 |
| MAE | 21.66% | 22.59% |

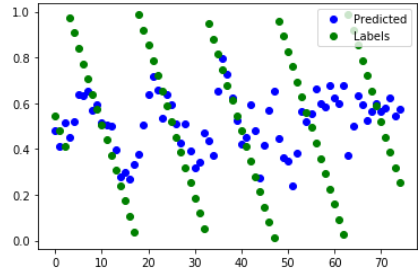

**Fig. 5.** The comparison of labels and predictions for the ChP-r model for the consequent 80 images

Despite the fact that the distribution of labels was uniform in the range from 0 to 1 and numerous experiments with different loss functions, the predictions of ChP-r models are inevitably concentrated in the middle of the time segment (0.5), which gives the average absolute error around 22-25% of the window slice width (see Fig. 5). This corresponds to approximately 16-19 business days for n_days = 75 or 5-7 business days for n_days = 25.

Same as for ChP-c, additional research is required to exclude label contradictions and calibrate the hyperparameters. A change in one hyperparameter may sufficiently change the result as it happened to the TF model, but this is the process of trial and error.

**TF model**. The dataset for the TF model uses the markup windows initially selected by the experts. Since the image is now built not by a slice of data of a fixed length, but by a markup window, the problem of duplicates and contradictions is not acute - it is unlikely that experts would select exactly the same windows, and, moreover, label them differently. The train and test data sets for TF models have around 10 thousand and 5 thousand of observations accordingly. In total, more than three dozen models were trained. However, the first twenty models consistently predicted the majority class (0 ("No trend, flat")), which automatically ensured accuracy around 60%. The training of some models with dpi=60 was interrupted due to the lack of RAM. Only after switching to the lower resolution and changing the learning rate α=0.001, a leap in quality was obtained. The results for dpi=20 and dpi=10 of the last 12 models were almost equal but it is preferable to use the TF model with dpi = 10 as less resource-intensive. The normalized confusion matrix of the best model (TF_19) is shown in Fig. **66**.

As we can see the model best determines flats (89% of correctly recognized observations of this class) and the uptrend (95%). The situation is relatively worse with the downward trend: in 25% of cases the TF_19 (dpi = 10) mistakenly recognizes it as flat. For reference, we note that the TF model with dpi = 20 recognizes flats and upwards a little better (93% and 92%, respectively), but in 37% of cases it mistakenly mistook a downward trend for a flat.

A selective visual analysis of observations with errors, however, showed that the initial markup was not always successful. This, to some extent, explains the non-decreasing share of errors (11-12%) in the test sample and, in general, allows us to consider the designed TF model quite accurate.

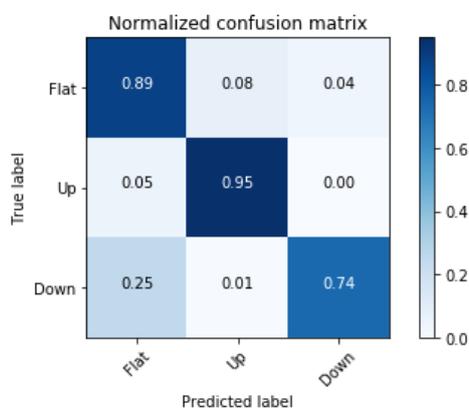

**Fig. 6.** The normalized confusion matrix for TF_19.

### 4.2 Simulation Results

The final stage of the proposed framework is a real-time simulation. This requires the development of quality metrics focused on financial results.

**Specific metrics**

Recalling that the direction of the trend can be determined by the slope of the regression line, we can calculate the profit earned during the time in position (i.e., during trends) and a couple of other metrics (see Table 4).

**Table 4.** The proposed quality metrics.

| Indicator | Description |
| --- | --- |
| Profit | The sum of all profits earned while in position (for all the stocks) |
| Days_in | The total number of business days in position (for all the stocks) |
| Times_in | The number of times the position was opened (for all the stocks) |
| DayProfit | The profit per one day in position, %: DayProfit= Profit/ Days_in |

| Indicator | Description |
|---|---|
| YearProfit | dayProfit scaled per annum, %: YearProfit= DayProfit*250, where 250 is the average number of business days in a year |
| YearProfit_avg | The average annual profit, including the days not in position, %: YearProfit_avg = Profit/number of data points *250 |

To compare models, the last two indicators - YearProfit and YearProfit_avg - are the most informative, because they are independent of the length of the time period and the number of stocks in the pipeline. From a business point of view, it might also be important to see how many times we opened the positions or what was the proportion between short or long for each stock, because it influences the additional costs of trading.

YearProfit shows how much profit, in percentage, one would gain on the money invested into assets if it was working during the whole year. But in reality, trends alternate with flat periods, so the money is not working all the time. YearProfit_avg shows profit including flat periods when no investments are made and of course will be smaller than YearProfit. To be more precise, we should, however, not consider flat periods as zero-profitable, since money can be put on overnight deposits. But for the purpose of this paper, we will treat the YearProfit_avg indicator just as a measure of effectiveness in identifying trading opportunities which results in bigger total profits for a chosen set of stocks and time period.

The main pitfall of YearProfit and YearProfit_avg metrics is that they are not normalized, so solely by their values we can't definitely say what is a good result. But we can compare our results with those, calculated on the same datasets labelled by real experts or even the virtual "average" expert. We should bear in mind, however, that experts were dealing with historical data, or "saw the future", so their profits are bound to be much and much higher.

**Results**. **Table 5** summarizes the results of trading simulation for n_days = 25 (ChP-c_4, ChP-r_15, TF_19 submodels) and n_days = 75 (ChP-c_6, ChP-r_13, TF_19 submodels). In both cases, the experiment was carried out for the entire test dataset (that is, data for dates between October 17, 2014, and May 13, 2017) for all 700 stocks. The skip step was set to 5 to reduce the number of iterations. However, even in this case, the running time of each simulation took about 16 hours on the 8 core 16 GB RAM Microsoft Azure virtual machine. The total number of data points for which the predictions were generated was 477 094 for n_days = 25 and 442 115 for n_days = 75.

**Table 5.** The simulation results.

| Metrics (short and long positions) | n_days=25 | | n_days=75 | |
|---|---|---|---|---|
| | Model | Expert | Model | Expert |
| Profit (%) | 812 | 14680 | 1859 | 11158 |
| Days_in | 101359 | 145293 | 130529 | 129325 |
| Times_in | 5088 | 832 | 4747 | 795 |
| YearProfit(%) | 2.00 | 25.26 | 3.56 | 21.57 |
| YearProfit_avg (%) | 0.43 | 7.69 | 1.05 | 6.3 |

The results of the trading simulation and their comparison with the results of the "average" expert indicate the need to refine the models. Even though positions are opened about 6 times more often (increases transaction costs), the final profit on them is several times less. To better understand the causes of errors, we analyzed the contingency matrices obtained as a result of the trading simulation.

For the case of n_days = 25, 63% of the days corresponding to upward trends and 86% of the days of downward trends are mistakenly recognized as flats, which does not allow making money on these trends. However, the model at least rarely confuses upward trends with downward trends, which would immediately lead to a loss.

In the case of n_days = 75 the final results are slightly higher. 85% of downwards are still recognized as flats, but the share of unrecognized upwards is 49%, which allows one to earn a little more on this trend in comparison with the case n_days = 25. Because most of the trends marked by experts are long-term and medium-term, we can conclude that using a wider data slice (n_days = 75) is more preferable to correctly identify the changepoints in market trends.

If we drill down into the P&L report of the trading simulation for n_days=75, we can notice, that it is short positions that cause losses. While, on average, the model lets us earn when we buy on the upward trend, it generates negative returns when we enter the market and sell at the beginning of the downward trend. This fact encourages us to refuse from short positions at all as non-efficient and concentrate only on traditional deals when one buys expecting the price will rise. Excluding short deals results in an increase in YearProfit_avg from 1.05% to 2.2% for n_days=75 case (see Table 6).

To decrease the number of unnecessary deals we can opt for a slightly different interpretation of the model signals. We will ignore flat signals which often are incorrect, that is, we will open the position at the first signal of the upward trend and close it only when we receive the first downward signal. This modification decreases the number of deals from 3365 to 1120, while YearProfit_avg reaches 4.76% and brings us closer to the "average expert" baseline - 5,83% (short positions excluded). Finally, we can compare our result with the buy-and-hold strategy, which assumes that we buy stock on the very first day and stay in position all the time. Surprisingly, it earns more than the "average expert" in terms of YearProfit_avg indicator - 9.22%, but uses money less effectively: the "average expert" stays in position 5 times less and therefore has the higher YearProfit indicator -24,8%. In conclusion we may say, that while the suggested model definitely allows to earn on increasing market it still does not outperform the alternatives, though some additional benefit may be gained from effectively using the time not in position, e.g. switching between the stocks in portfolio.

**Table 6.** The simulation results (short position excluded)

| Metrics | n_days=75 | | | |
|---|---|---|---|---|
| (short positions excluded) | Model | Model (flat ignored) | Expert | Buy and hold |
| Profit (%) | 3928 | 8414 | 10324 | 16302 |
| Days_in | 117723 | 369128 | 104056 | 442115 |
| Times_in | 3365 | 1120 | 655 | 699 |
| YearProfit(%) | 8.34 | 5.7 | 24.8 | 9.22 |

| YearProfit_avg (%) | 2.22 | 4.75 | 5.83 | 9.22 |

## 5 Conclusions

The paper illustrates the application of CNNs, traditionally used for image detection and recognition, to the problem of long-term market trend prediction. Unlike in the traditional approaches, the labels (trend or flat) are not derived from prices but filled manually by experts who worked with stock data as with images. The task is quite challenging since we actually try to 'digitalize' successful traders' skills and we can only compare the performance of the model with the performance of the experts themselves. The comparison with the results of other researchers would be inadequate since we use a completely different source for ground truth.

The main reason for the unsatisfactory accuracy of the predictions is the shortcomings of the ChP-c and ChP-r submodels described in the sections above. They, in turn, can be caused by inconsistencies in the original data markup. We should also consider the imbalance in the class of downward trends, which led to a relatively lower accuracy of their recognition by the TF model. In general, however, the proposed CNN framework for the prediction of changepoints in long-term market trends allows us to learn from successful traders' decisions and evaluate the model performance, although the submodels require additional exploring of errors and calibration.

Improving the prediction quality of the ChP-c, ChP-r and TF models is the main direction of further research. One of the options is to use a pretrained network (for example, Alexnet [16] or YOLO [17]) and adapt it for ChP-c and ChP-r labels by changing the fully connected output layer. We also consider a radically different model design with gradient boosting algorithm XGBoost, which currently performs comparatively better, but too, requires additional research.

The most successful trading simulation results are shown in Fig. 7 and Fig. 8 as an illustration.

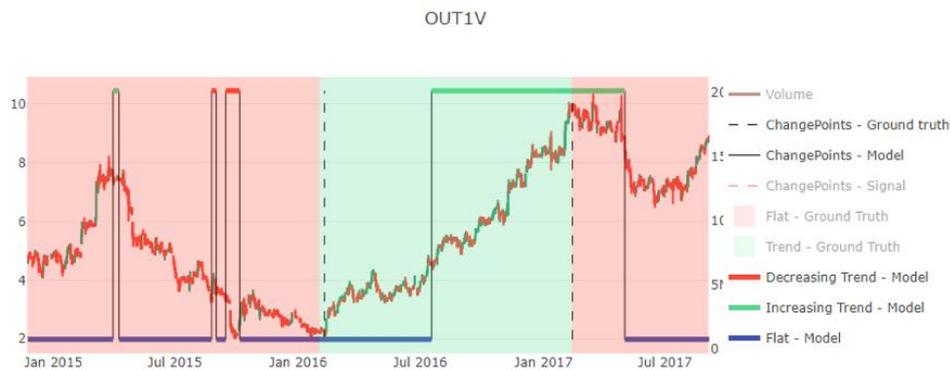

**Fig. 7.** An example of simulation for n_days = 25

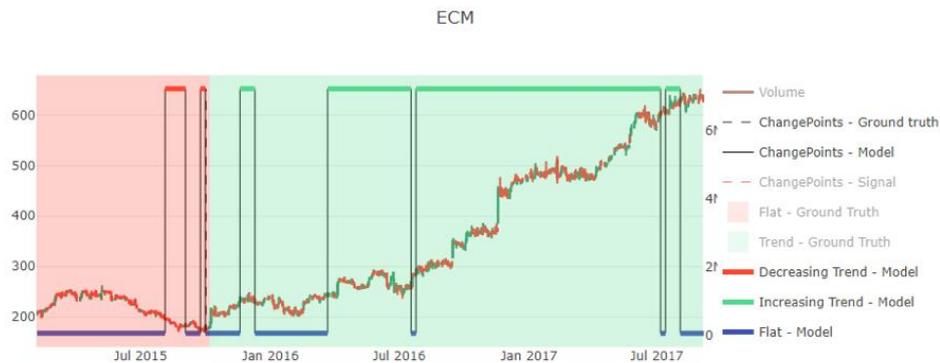

**Fig. 8.** An example of simulation for n_days = 75